\documentstyle[12pt]{article}
\newcommand{\matel}[3]{\langle #1|#2|#3\rangle}
\newcommand{\ra}{\rightarrow}
\newcommand{\sG}{\sigma \cdot G}
\newcommand{\aver}[1]{\langle #1\rangle} 

\newcommand{\as}{\alpha_s}
\begin{document}
\begin{flushright} 
UND-HEP-96-BIG06\\
November 1996\\
\end{flushright}
\vspace*{.3cm}  
\begin{center}
\begin{Large}    
{\bf HEAVY QUARK EXPANSIONS FOR INCLUSIVE HEAVY-FLAVOUR  
DECAYS AND THE LIFETIMES OF CHARM AND BEAUTY HADRONS}
\footnote{Invited Lecture given at HQ96 "Heavy Quarks at 
Fixed Target", St. Goar, Germany, Oct. 3 - 6, 1996}  \\ 
\end{Large} 
\vspace*{.3cm}
Ikaros I. Bigi  \\
\vspace{.4cm}
{\em Physics Depart., University of Notre Dame du Lac, 
Notre Dame, IN 46556, U.S.A.} \\
{\em e-mail:BIGI@UNDHEP.HEP.ND.EDU} \\ 
\vspace*{.4cm}
{\bf ABSTRACT}
\end{center} 
Inclusive heavy-flavour decays can be described through $1/m_Q$ 
expansions derived from QCD with the help of an operator product 
expansion. I sketch their methodology and apply them first to 
semileptonic $B$ decays; $|V(cb)|$ can be extracted from 
$\Gamma _{SL}(B)$ and $\bar B \ra l \nu D^*$ with the result: 
$|V(cb)|_{incl} = 0.0413 \pm 0.0016_{exp} \pm 0.002_{theor}$, 
$|V(cb)|_{excl} = 0.0377 \pm 0.0016_{exp} \pm 0.002_{theor}$. 
The lifetimes of charm and beauty hadrons are discussed. The 
charm lifetimes are predicted/reproduced as well as could be 
expected. Predictions on $B$ meson lifetimes agree with available 
data; $\Lambda _b$ baryons are predicted to be shorter lived 
than $B_d$ mesons by no more than $\sim$ 10 \% -- in marked 
contrast to present measurements. I evaluate the situation and 
comment on recent theoretical criticism. The importance of the 
concepts of global vs. local quark-hadron duality is pointed out.  
\baselineskip=17pt
\vspace*{0.5cm} 
\section{Motivations}
Although I am certainly preaching to the converted here, I want to 
start out with some very general statements on the virtues of 
heavy-flavour physics. Analysing it in detail 
plays a central role in our attempts to 
uncover Nature's Grand Design. For this dynamical sector provides us 
with an extremely rich phenomenology possessing fundamental 
implications with CP violation as the ultimate prize. 
Few other (if any) areas have a comparable potential for 
fundamental discoveries. 
Some predictions can be made with high 
{\em parametric} reliability; for example the CP asymmetry in 
$B_d \rightarrow \psi K_S$ can be predicted to be 
given by 
sin$2\beta$ with very high reliability. Furthermore new theoretical 
tools apply enabling us to translate the {\em parametric} 
into a {\em numerical} accuracy. This will be the subject of my 
talk here. 

As explained later I find it quite realistic that the numerical values 
of the KM angles of most direct relevance for beauty physics can be 
extracted with the following accuracy: 
\begin{eqnarray}
\delta |V(cb)| &\leq &\; \pm \,  
{\rm very\; few\;  } \% 
\label{goal1}  
\\
\delta \left| \frac{V(ub)}{V(cb)}\right| &\leq &\; \pm \,  
{\rm few\;  } \% 
\label{goal2} 
\\ 
\delta \left| \frac{V(td)}{V(cb)}\right| &\leq &\; \pm \, 
(10 - 15) \, \%
\label{goal3} 
\end{eqnarray}
The objective set in eq.(\ref{goal1}) is a near-term goal we 
are close to achieving; eqs.(\ref{goal2}) and (\ref{goal3}) amount to  
mid-term and long-term goals, respectively. 

These benchmarks represent tall orders; our hope to attain them 
is based on the presence of the heavy-flavour quark mass: 
expanding transition amplitudes in powers of $\mu /m_Q$ 
-- with $\mu $ denoting an ordinary hadronic scale 
not exceeding 1 GeV -- should lead to meaningful results for 
beauty decays when only the first few terms are retained 
since $\mu /m_Q \ll 1$; for charm 
on the other hand, the situation is a priori unclear. 

This hope can be formulated through four second-generation 
theoretical technologies, namely (i) QCD sum rules, 
(ii) Lattice QCD, (iii) Heavy Quark Effective Theory (=HQET) 
\cite{MANNEL} 
and (iv) Heavy Quark Expansions; methods (i) - (iii) deal 
primarily with exclusive decays whereas (iv) treats inclusive 
transitions \footnote{It will be pointed out later that methods 
(iii) and (iv) are distinct and should not be equated.}.  
The message I want to convey is the following: 
\begin{itemize} 
\item 
There are several self-consistent methods that 
to the best of our knowledge are 
genuinely based on QCD.  
\item 
They allow a {\em quantitative} treatment of important aspects of 
the decays of heavy-flavour hadrons. 
\item 
Even failures teach us important lessons on QCD, however saddening 
they might be in that case. 
\end{itemize} 
The remainder of my talk will be organized as follows: in Sect. 2 I 
introduce the theoretical tools, which in Sect. 3 are applied to 
semileptonic beauty decays; in Sect. 4 I discuss the lifetimes of 
charm and beauty hadrons before presenting my 
conclusions in Sect. 5. 

\section{The $1/m_Q$ Methodology for Inclusive Decays 
of Heavy-Flavour Hadrons}
\subsection{The Operator Product Expansion}
Analogous to the treatment of 
$e^+e^-\rightarrow hadrons$ one can describe the decay rate 
into an 
inclusive final state $f$ in terms of the imaginary part of a 
forward scattering operator evaluated to second order in the 
weak interactions \cite{SV,BUV,BS}:
\begin{equation} 
\hat T(Q\rightarrow f\rightarrow Q)=
i \, {\rm Im}\, \int d^4x\{ {\cal L}_W(x){\cal L}_W^{\dagger}(0)\} 
_T
\label{OPTICAL} 
\end{equation} 
where $\{ .\} _T$ denotes the time ordered product and 
${\cal L}_W$ the relevant effective weak Lagrangian expressed on 
the 
parton level. If the energy released in the decay is sufficiently 
large 
one can express the {\em non-local} operator product in 
eq. (\ref{OPTICAL}) 
as an 
infinite sum of {\em local} operators $O_i$ of increasing 
dimension 
with 
coefficients $\tilde c_i$ 
containing higher and higher powers of  
$1/m_Q$\footnote{It should be kept in mind, 
though, that 
it is primarily the {\em energy release} rather than $m_Q$ that 
controls the 
expansion.}. {\em This operator product expansion (OPE) 
\cite{WILSON} and its 
consistent realization is the central theoretical tool in the Heavy 
Quark Expansions.} The width for $H_Q\rightarrow f$ is then 
obtained by 
taking the 
expectation value of $\hat T$ for the heavy-flavour hadron $H_Q$:
\begin{equation} 
\matel{H_Q}{\hat T (Q\rightarrow f\ra Q)}{H_Q} \propto 
\Gamma (H_Q\rightarrow f) = G_F^2 |KM|^2 
\sum _i \tilde c_i^{(f)}(\mu ) 
\matel{H_Q}{O_i}{H_Q}_{(\mu )}\; ,   
\label{OPE}
\end{equation} 
where I have used the following notation: $|KM|$ denotes the 
appropriate combination of KM parameters; the 
c-number coefficients $\tilde c_i^{(f)}(\mu )$ are 
determined by short-distance dynamics whereas long-distance 
dynamics control the expectation values of the local 
operators $O_i$. Such a separation necessitates the introduction of 
an auxiliary scale with {\em long distance $>\mu ^{-1} >$   
short distance} \cite{WILSON}. While this is a conceptually and often 
also practically important point I will not refer to it 
explicitely anymore in this article 
\footnote{Observables of course do not depend on $\mu$. 
Yet one has to choose $\Lambda _{QCD} \ll \mu \ll m_Q$ 
if one wants to calculate perturbative as well as non-perturbative 
corrections in a self-consistent way.}. 

The master formula eq.(\ref{OPE}) holds for a host of different  
integrated 
heavy-flavour decays: semileptonic, nonleptonic and radiative 
transitions, 
KM favoured or suppressed etc. For semileptonic and nonleptonic 
decays, treated to order $1/m_Q^3$, it takes the 
following form \footnote{Expanding
$\langle H_Q|\bar Q i \sigma \cdot G Q|H_Q\rangle /m_Q^2$
also yields contributions of order $1/m_Q^3$; those are however
practically insensitive to the light quark flavours.}:  
$$\Gamma (H_Q\ra f)=\frac{G_F^2m_Q^5}{192\pi ^3}|KM|^2
\left[ c_3^f\matel{H_Q}{\bar QQ}{H_Q}+
c_5^f\frac{
\matel{H_Q}{\bar Qi\sG Q}{H_Q}}{m_Q^2}+ \right. 
$$
\begin{equation}  
\left. +\sum _i c_{6,i}^f\frac{\matel{H_Q}
{(\bar Q\Gamma _iq)(\bar q\Gamma _iQ)}{H_Q}}
{m_Q^3} + {\cal O}(1/m_Q^4)\right]  
\label{WIDTH} 
\end{equation} 
Four comments might elucidate this expression: 
\begin{itemize}
\item 
We know which local operators can appear in the operator 
product expansion and what their dimensions are; this determines 
how they scale with $m_Q$. We also follow the usual 
procedure of actually calculating the short-distance coefficients 
$c_i(f)$ in perturbation theory. 
\item 
The expectation values of the local operators are shaped by 
long-distance dynamics and in general we cannot derive their 
size from first principles. We will employ $1/m_Q$ expansions 
to relate these matrix elements to other observables of a typically 
static nature like hadron masses. One can also rely on the findings 
from QCD sum rules and lattice QCD concerning these expectation 
values; this will be of increasing value in the future. 
\item 
Eq.(\ref{WIDTH}) does {\em not} contain a contribution of order 
$1/m_Q$ since there is {\em no independant gauge-invariant 
dimension-four} 
operator \footnote{The situation is more subtle for final-state 
spectra \cite{PRL}.}. The leading non-perturbative corrections are 
then 
of order $(\mu /m_Q)^2$. As we will see in more detail this 
means they amount to no more than a few percent in beauty decays. 
This is one major reason why the hope to achieve the benchmark 
accuracies listed above is realistic: one has to control the 
non-perturbative corrections only on the, say, 30\% level to 
describe an integrated witdth with a few percent accuracy. 
\item 
{\em The expansion parameter is actually the inverse of the 
energy release rather than $1/m_Q$} although this is not manifest 
in the expression given in eq.(\ref{WIDTH}). This distinction will 
become relevant for the discussion of $b \ra c \bar c s$. 
\end{itemize} 
The three terms appearing on the right-hand side of 
eq.(\ref{WIDTH}) allow an intuitive interpretation: (i) The leading 
operator $\bar QQ$ contains the spectator contribution 
that dominates for $m_Q \ra \infty$, yet goes beyond it: for example 
in incorporates the motion of the heavy quark relative to the 
rest frame of the hadron. 
(ii) $\matel{H_Q}{\bar Q i\sG Q}{H_Q}$ describes the spin interaction 
of the heavy quark $Q$ with the light degrees of freedom inside the 
hadron. This term had been overlooked in the 
earlier phenomenological descriptions. 
(iii) $\matel{H_Q}
{(\bar Q\Gamma _iq)(\bar q\Gamma _iQ)}{H_Q}$ contains the 
so-called Pauli Interference (PI) 
\cite{PI}, Weak Annihilation (WA) \cite{WEX} and 
Weak Scattering (WS) \cite{WS} contributions which had been introduced by 
the earlier phenomenological descriptions. However there is little 
`wiggle space' here: WA is helicity suppressed 
\cite{MIRAGE}; PI and WS scale 
(at least formally) like $1/m_Q^3$. 
\subsection{Determination of the Expectation Values}
Using the equation of motion one can expand the local operator 
$\bar QQ$ in powers of $1/m_Q$ and finds 
\footnote{I use here a relativistic normalization: 
$\matel{H_Q}{O_i}{H_Q}_{norm} \equiv 
\matel{H_Q}{O_i}{H_Q}/2M(H_Q)$. }: 
\begin{equation} 
\matel{H_Q}{\bar QQ}{H_Q}_{norm} = 1 + 
\frac{\matel{H_Q}{\bar Q\frac{i}{2}\sG Q}{H_Q}_{norm}}{2m_Q^2} 
- \frac{\aver{(\vec p_Q)^2}_{H_Q}}{2m_Q^2} + {\cal O}(1/m_Q^3) 
\label{QBARQ}
\end{equation}
The first term on the right-hand side, which reflects the flavour 
charge carried by $H_Q$, represents the spectator contribution that 
dominates for 
$m_Q \ra \infty$. 

\noindent The expectation values of the chromomagnetic 
operator are known. Since the light di-quark system inside 
$\Lambda _Q$ and $\Xi _Q$ (but not inside 
$\Omega _Q$) baryons carries no spin, there can be no 
spin-interaction: 
\begin{equation} 
\matel{\Lambda _Q}{\bar Q\frac{i}{2}\sG Q}{\Lambda _Q} \simeq 0 
\simeq \matel{\Xi _Q}{\bar Q\frac{i}{2}\sG Q}{\Xi _Q} 
\label{SGLAMBDA}
\end{equation}
The expectation value for pseudoscalar mesons $P_Q$ is given by the 
observed hyperfine splitting between the masses of the vector $V_Q$  
and pseudoscalar mesons: 
\begin{equation}
\matel{P_Q}{\bar Q\frac{i}{2}\sG Q}{P_Q}_{norm} \simeq 
\frac{3}{4}\left( M^2(V_Q) - M^2(P_Q)\right) 
\label{HYPERFINE} 
\end{equation} 
For beauty and charm one then has 
\begin{eqnarray}
G_b &\equiv & 
\frac{\matel{B}{\bar b\frac{i}{2}\sG b}{B}_{norm}}{m_b^2} 
\simeq 0.016 
\label{GB}
\\ 
 G_c &\equiv & 
\frac{\matel{D}{\bar c\frac{i}{2}\sG c}{D}_{norm}}{m_c^2} 
\simeq 0.21 
\label{GC} 
\end{eqnarray} 
This representing a second order correction one infers that the 
expansion parameter is small albeit not tiny for beauty decays:  
$\sqrt{G_b} \sim 0.13$; for charm it is not small though at least 
smaller than unity: $\sqrt{G_c} \sim 0.46$. 

The expectation values of the other independant dimension-five 
operator 
\begin{equation} 
\matel{H_Q}{\bar Q (i\vec D)^2Q}{H_Q}_{norm} 
\equiv \aver{(\vec p_Q)^2}_{H_Q}  
\label{KINEN} 
\end{equation}
-- with $D_{\mu}$  denoting 
the covariant derivative -- can be interpreted as the average 
kinetic energy of the heavy quark $Q$ inside the hadron 
$H_Q$. Its numerical value is not known precisely. From an analysis 
based on QCD sum rules \cite{QCDSR} one obtains  
\begin{equation} 
 \aver{(\vec p_b)^2}_B \simeq 0.5 \pm 0.1 \; ({\rm GeV})^2 
\label{BALLBRAUN} 
\end{equation} 
in agreement with a lower bound \cite{OPTICAL,VOLOSHINBOUND} 
\begin{equation} 
\aver{(\vec p_b)^2}_B \geq 
\matel{B}{\bar b\frac{i}{2}\sG b}{B}_{norm} \pm 
0.15 \; ({\rm GeV})^2 = (0.37 \pm 0.15)\; ({\rm GeV})^2 
\label{BOUND1} 
\end{equation} 
The differences in the mesonic and baryonic expectation values 
can be related to the `spin averaged' meson and baryon masses: 
$ \langle (\vec p_Q)^2\rangle _{\Lambda _Q}-
\langle (\vec p_Q)^2\rangle _{P_Q} \simeq 
\frac{2m_bm_c}{m_b-m_c}\cdot 
\{ [\langle M_D\rangle -M_{\Lambda _c}]- 
[\langle M_B\rangle -M_{\Lambda _b}] \}$ \cite{BUVPREPRINT}. 
Present data yield:  
\begin{equation} 
\langle (\vec p_Q)^2\rangle _{\Lambda _Q}-
\langle (\vec p_Q)^2\rangle _{P_Q} =  
- (0.015 \pm 0.030)\; {\rm (GeV)^2} 
\label{KINENBVM}
\end{equation} 
i.e., no significant difference.  
In deriving eq.(\ref{KINENBVM}) it was assumed that 
the $c$ quark can be treated as heavy; in that case  
$\langle (\vec p_c)^2\rangle _{H_c} \simeq 
\langle (\vec p_b)^2\rangle _{H_b}$ holds. 

The expectation values of the two classes of four-fermion operators 
that appear -- one coupling two colour singlets, the other two 
colour octets -- are not known accurately. To estimate their size for 
{\em mesons} 
one usually invokes {\em factorization} or vacuum saturation:   
$$ 
\matel{P_Q(p)}{J_{\mu}^{(0)}\cdot J_{\nu}^{(0)}}{P_Q(p)} \equiv  
\matel{P_Q(p)}{(\bar Q_L\gamma _{\mu}q_L)
(\bar q_L\gamma _{\nu}Q_L}{P_Q(p)}_{norm}\simeq 
$$
\begin{equation} 
\matel{P_Q(p)}{(\bar Q_L\gamma _{\mu}q_L)}{0}_{norm} 
\matel{0}{(\bar q_L\gamma _{\nu}Q_L}{P_Q(p)}_{norm} = 
\frac{1}{8 M_{P_Q}}f^2_{P_Q}p_{\mu}p_{\nu}
\label{FACT2SS} 
\end{equation}    
$$
\matel{P_Q(p)}{J_{\mu}^{(i)}\cdot J_{\nu}^{(i)}}{P_Q(p)} \equiv  
\matel{P_Q(p)}{(\bar Q_L\gamma _{\mu}\lambda _iq_L)
(\bar q_L\gamma _{\nu}\lambda _iQ_L}{P_Q(p)}_{norm}\simeq 
$$
\begin{equation} 
\matel{P_Q(p)}{(\bar Q_L\gamma _{\mu}\lambda _i q_L)}{0}_{norm} 
\matel{0}{(\bar q_L\gamma _{\nu}\lambda _i Q_L}{P_Q(p)}_{norm} 
= 0 
\label{FACT2OO} 
\end{equation}    
Such an ansatz cannot be an identity; it can hold as an 
approximation, though -- at  
certain scales. Invoking it at $\sim m_Q$ does {\em not} 
make sense at all. For as far as QCD is concerned, $m_Q$ is a 
completely 
foreign quantity. A priori it has a chance to hold at ordinary hadronic 
scales 
$\mu  \sim 0.5 \div 1$ GeV \cite{HYBRID}; 
various theoretical analyses based on 
QCD sum rules, QCD lattice simulations, $1/N_C$ expansions etc. have 
indeed found it to apply in that regime. It would be  
inadequate conceptually as well as numerically to renormalize 
merely the decay constant: $f_Q(m_Q) \ra f_Q(\mu )$. Instead 
the 
full set of operators has to be evaluated at $\mu $. One 
proceeds 
in three steps (for details see \cite{BELLINI}): 

\noindent (A) Ultraviolet renormalization translates 
the weak Lagrangian defined at $M_W$, ${\cal L}_W(M_W)$, into 
one effective at $m_Q$, ${\cal L}_W(m_Q)$. 

\noindent (B) The operators $J_{\mu}^{(0)}\cdot J_{\nu}^{(0)}$ and   
$J_{\mu}^{(i)}\cdot J_{\nu}^{(i)}$ undergo hybrid 
renormalization \cite{HYBRID} down to $\mu $. 

\noindent (C) At scale $\mu $ one invokes factorization. 

\noindent To be more specific I state the final result for the PI 
contribution: 
\begin{equation} 
\Delta \Gamma _{PI}\simeq \Gamma _0\cdot 
24\pi ^2\frac{f_{H_Q}^2}{M_{H_Q}^2}\kappa ^{-4} 
\left[ (c_+^2-c_-^2)\kappa ^{9/2}+\frac{c_+^2+c_-^2}{3} 
-\frac{1}{9}(\kappa ^{9/2}-1)(c_+^2-c_-^2)\right] 
\label{PI} 
\end{equation}  
where $\Gamma _0$ denotes the width for the decay of a free 
quark $Q$ and $c_{\pm}$ the usual UV operator renormalization 
coefficients; hybrid renormalization is described by  
\begin{equation} 
\kappa \equiv \left[ \frac{\as (\mu ^2)}
{\as (m_Q^2)}\right] ^{1/b}, \; 
b=11-\frac{2}{3} n_F  
\label{KAPPA}  
\end{equation} 
From eq.(\ref{PI}) one reads off for the colour factor 
\begin{equation} 
\left[ (c_+^2-c_-^2)\kappa ^{1/2}+\frac{c_+^2+c_-^2}{3\kappa ^4} 
-\frac{(\kappa ^{9/2}-1)}{9\kappa ^4}(c_+^2-c_-^2)\right] 
\longrightarrow \left[ \frac{4}{3} c_+^2 - \frac{2}{3}c_-^2\right]  
\longrightarrow \frac{2}{3} 
\end{equation}
when first ignoring hybrid renormalization -- $\kappa =1$ -- 
and then UV renormalization as well -- $c_+=1=c_-$. 

\noindent Some comments are in order for proper evaluation: 
\begin{itemize} 
\item 
The 
factorizable contributions to PI largely cancel at scales around $m_Q$ 
-- in particular in the case of beauty and 
apparently for accidental reasons. 
The ratio of non-factorizable to factorizable contributions is thus 
large and numerically unstable there. 
\item No such cancellation occurs around scales $\mu _{had}$ 
making  
factorizable contributions numerically more stable and dominant 
over 
non-factorizable ones. 
\item Contributions that are factorizable (in colour space) 
at $\mu _{had}$ are mainly 
non-factorizable at $m_Q$. 
\item The role of 
non-factorizable terms has been addressed in the literature over the 
years, most explicitely and in a most detailed way in \cite{DS,WA}. 
\end{itemize}
I will later comment on the criticism expressed in \cite{NS}. 

The situation becomes much more complex for  
baryon decays. To order $1/m_Q^3$ there are several different 
ways in which the valence quarks of the baryon can be contracted 
with the quark fields in the four-quark operators; furthermore 
WS is {\em not} helicity suppressed and thus can make a sizeable 
contribution to lifetime differences; also the PI effects can 
now be constructive 
as well as destructive. Finally one cannot take 
recourse to factorisation as a limiting case.  
Thus there emerge three types of 
numerically significant mechanisms at this order in baryon 
decays -- in contrast to meson decays where there is a 
{\em single}  dominant 
source for lifetime differences -- and their strength cannot be 
expressed in terms of 
a single observable like $f_{H_Q}$. At present we do not know 
how to determine the relevant matrix elements in a 
model-independant way. Guidance and 
inspiration has traditionally been derived from quark model calculations with 
their inherent uncertainties. This analysis had already been 
undertaken 
in the framework of phenomenological models 
\cite{BARYONS1,BARYONS2,BARYONS3}. 
One thing should be obvious already at this point: 
with terms of different signs and somewhat uncertain size 
contributing to differences among baryon lifetimes one has to 
take even semi-quantitative predictions with 
a grain of salt! 

There is another relation that will become highly relevant in the 
discussion of semileptonic beauty decays. The mass difference 
which is free of renormalon ambiguities and well-defined can be 
expressed as follows:
\begin{equation} 
m_b - m_c \simeq  \frac{1}{4}(M_B + 3 m_{B^*}) - 
\frac{1}{4}(M_D + 3 m_{D^*}) + 
\aver{(\vec p_Q)^2}_{P_Q}\cdot \left( \frac{1}{2m_c} - 
\frac{1}{2m_b} \right) 
\label{MBMC} 
\end{equation} 
Using the {\em observed} mass values for the charm and beauty 
mesons and varying $\aver{(\vec p_Q)^2}_{P_Q}\simeq 
\aver{(\vec p_b)^2}_B\simeq \aver{(\vec p_c)^2}_D$ over a 
{\em reasonable} range one obtains 
\begin{equation} 
m_b - m_c \simeq (3.46 \pm 0.04) \; {\rm GeV} 
\label{MBMCNUM} 
\end{equation}
\section{Semileptonic $B$ Decays}
There are three topics I want to address here, namely the 
semileptonic branching ratio of $B$ mesons and the extractions 
of $|V(cb)|$ and $|V(ub)|$. 
\subsection{$BR_{SL}(B)$}
Without radiative QCD corrections one finds 
$BR(b \ra l \nu c)\sim 15\%$. Including them lowers 
$BR(b \ra l \nu c)$ down to $\sim 12 - 13.5\%$. Yet this is still 
significantly higher than the observed branching ratio for beauty 
{\em mesons} \cite{PDG96}: 
\begin{equation} 
BR(B \ra l \nu X) = 10.43 \pm 0.24 \% 
\label{BRSLDATA} 
\end{equation} 
The weak link in the theoretical treatment is the estimate of 
$\Gamma (\bar B \ra c \bar c s \bar q)$ since the energy release in 
$b \ra c \bar c s$ is not very large; therefore the nonperturbative 
corrections might not be under good control. It is then quite 
conceivable that $BR(\bar B \ra c \bar c s \bar q)$ is 
considerably larger than is usually inferred from 
$BR(b \ra c \bar c s)$ computed on the parton level; for 
this latter quantity a `canonical' value of 0.15 is often 
adopted 
\footnote{Being `canonical' does not make it necessarily 
right -- even for someone from Notre Dame.}. 
There are actually various 
{\em theoretical} indications that an enhancement of this 
nonleptonic channel indeed 
takes place \cite{GCCBARS}. 
One can then entertain the idea that 
$BR(\bar B \ra c \bar cs \bar q) \sim 0.3$ holds rather than $0.15$;  
this would bring the predicted semileptonic 
branching ratio into agreement with the observed one.
One has to keep in mind that 
this transition rate is particularly sensitive to which values one 
adopts for the quark masses: using `low' values for 
the quark masses, namely $m_b=4.6$ GeV, $m_c=1.2$ GeV 
and $m_s=0.15$ GeV, rather than `high' values -- 
$m_b=5.0$ GeV, $m_c=1.7$ GeV 
and $m_s=0.30$ GeV-- would enhance 
$BR(b \ra c \bar cs)$ by a factor of about two to a value 1.23 
\cite{PETRARCA,STONEFIRST} 
\footnote{One should note that both sets of mass values 
satisfy $m_b - m_c \simeq 3.4$ GeV.}! This magnified 
dependance on the quark mass values provides us with 
another illustration that 
nonperturbative corrections can be expected to be large here 
since they control the proper usage of quark masses.  It might 
still turn out that we will be able to calculate 
$\Gamma (\bar B \ra c \bar c s \bar q)$ reliably -- once we 
understand how the quark masses are to be evaluated for 
this transition.    

Such a resolution of the puzzle would have another observable 
consequence: it would lead to a larger than previously expected 
charm yield in $B$ decays. To be more specific: for $n_c$ -- 
the number of charm states emerging from $B$ decays -- 
one has $n_c \simeq 1 +BR(\bar B \ra c \bar c s \bar q)$. 
This quantity 
can be measured where one assigns charm multiplicity {\em one} 
to $D$, $D_s$, $\Lambda _c$ and $\Xi _c$ and {\em two} to 
charmonia. There are two new experimental studies which I 
juxtapose in Table \ref{RESULTSNC} to the two theoretical 
expectations sketched above: 
\begin{table}
\centering
\caption{ \it The measured values of $n_c$.
}
\vskip 0.1 in
\begin{tabular}{|l|c|c|} \hline
          &   CLEO 
&ALEPH  \\
\hline
\hline
 $n_c$  =      & $1.134 \pm 0.043$& $1.23 \pm 0.07$ 
 \\
\hline
\end{tabular}
\label{RESULTSNC}
\end{table}
While both experimental numbers \cite{CASSEL} are consistent with each other and 
the `canonical' value 1.15, the 
CLEO number clearly favours 1.15 over 1.3. Yet in view of the ALEPH 
number one can say that a higher value of 1.25 - 1.30 that would 
lead to predicting the observed semileptonic branching ratio is not 
ruled out yet. 

The plot thickens even further: the channel 
$\bar B \ra c \bar c s \bar q$ can be accessed more directly by  
observing the decays of $B$ into `wrong-sign' $D$ mesons 
as done by CLEO: 
\begin{equation} 
BR(\bar B = (b \bar q) \ra \bar D=(c\bar q) + X) = 
0.081 \pm 0.026 
\label{WRONGSIGN} 
\end{equation} 
Such an inclusive transition is fed almost completely by 
$b\ra c \bar cs$ since the strong fragmentation reaction 
$q \ra q c \bar c$ is highly suppressed. Combining 
eq.(\ref{WRONGSIGN}) with the findings on 
$\bar B\ra \bar D_s +X$, $\psi ^{(\prime )} +X$, etc. yields 
\begin{equation} 
BR(\bar B \ra c \bar c s \bar q) = 0.239 \pm 0.038 
\label{BCCBARS} 
\end{equation} 
The measurement given in eq.(\ref{WRONGSIGN}) means that the 
channel $\bar B\ra c \bar c s \bar q$ is 
{\em not} dominated by 
$\bar B \ra \bar D_s+X$ --  
contrary to earlier expectations! It should be noted that a recent 
theoretical analysis \cite{BSU} invoking factorization finds 
$\Gamma (\bar B \ra \bar D_s+X)\leq 
\frac{1}{2}\Gamma (\bar B \ra c \bar c s \bar q)$ due to the 
production of higher-mass $\bar D_s^{*... }$ resonances decaying 
preferentially into $\bar D + K +X$. 

My conclusions  are the following: The experimental situation is still 
somewhat in limbo 
with various intriguing possibilities. The present discrepancy 
between the data and the theoretical expectation will probably be 
resolved through a combination of factors. One also has to allow 
for a larger theoretical uncertainty in predicting the {\em absolute} 
value of this branching ratio than the ratio of semileptonic 
branching ratios. 

\subsection{Extracting $|V(cb)|$} 
There is near-universal consensus in the community that the KM 
parameter $|V(cb)|$ is best extracted from semileptonic $B$ 
decays (and likewise for $|V(ub)|$). This consensus gets 
dissipated, however, when one discusses what specifically is the most 
reliable method for that. I will sketch two complementary 
analyses which I consider the most reliable ones from a theoretical 
perspective. 
\subsubsection{$|V(cb)|$ from $\Gamma _{SL}(B)$} 
The total semileptonic width of $B$ mesons being proportional to 
$|V(cb)|^2$ is a prime candidate: 
\begin{equation} 
\Gamma _{SL}(B) = \frac{G_F^2m_b^5}{192 \pi ^3} |V(cb)|^2 \cdot 
\left[ F(\as , m_c^2/m_b^2, \aver{(\vec p_b)^2}_B/m_b^2) 
+ {\cal O}(\as ^2, \as /m_b^2, 1/m_b^3)\right]  
\label{GAMMASL} 
\end{equation}
The function $F$ in eq.(\ref{GAMMASL}) 
containing perturbative, mass and nonperturbative 
corrections \cite{BUV} is known;  
in general nonperturbative 
corrections are found to be small and under control. Yet the 
dependance on the fifth power of the $b$ quark mass with its 
intrinsic uncertainties would appear 
-- at first sight -- to severely limit or even vitiate the quantitative 
usefulness of $\Gamma _{SL}(B)$. However it turns out that 
$\Gamma _{SL}(B)$ depends mainly on the difference 
$m_b - m_c$ rather than on $m_b$ and $m_c$ separately 
although this is not manifest in eq.(\ref{GAMMASL}). It is 
obvious in the 
so-called Small Velocity (SV) limit that is realized for 
$(m_b - m_c)/m_b \ll 1$, i.e. $m_c \simeq m_b$; it is shown in \cite{FIVE} 
that there is an extended SV limit, i.e. the 
SV limit becomes relevant precociously for the real 
value of $m_c/m_b \sim 1/3$.  As pointed out above, the 
difference $m_b - m_c$ is well-defined and well-known 
numerically to within 1 \% roughly, see eq.(\ref{MBMCNUM}). 

The new criticism was put forward that the 
{\em perturbative} expansion is particularly ill-behaved, namely 
that the coefficient for the second order term $(\as /\pi )^2$ is 
around 10 for $b\ra c$ and even 20 for $b \ra u$ 
\cite{WISE}. Yet 
considerable care has to be applied in treating quark masses.   
Usage of the {\em pole} mass is not appropriate when including 
perturbative as well as nonperturbative corrections 
\cite{RENORMALON}; 
{\em running}  
masses on the other hand can be employed.   
When extracting $m_b$ from $\Upsilon$ spectroscopy 
\cite{MBVOL} 
and applying it to $\Gamma _{SL}(B)$ one has to keep track 
at which scale $m_b$ is evaluated. Using $m_b$ evaluated at 
the high scale $m_b$ indeed leads to the appearance of large 
second order contributions in $\Upsilon$ spectroscopy 
as well as in $B$ decays! However, once one evaluates $m_b$ at a 
low scale around 1 GeV, the coefficients of the second order 
corrections become small, namely less than unity 
\cite{VCBPERT}. It suggests that the 
natural scale is not the heavy mass, but considerably 
smaller. This observation can be explained through a careful analysis 
of the phase space available in semileptonic $B$ decays 
\cite{FIVE}. 

Putting everything together one arrives at 
\begin{equation} 
|V(cb)|_{incl} = (0.0413 \pm 0.002_{theor})\times 
\sqrt{\frac{1.57\; {\rm psec}}{\tau (B)}} \cdot 
\sqrt{\frac{BR_{SL}(B)}{0.1043}} 
\label{VCBIN}
\end{equation} 
where I have listed the theoretical uncertainty only, estimated to 
be around 5\% based on the following considerations: 
\begin{enumerate}
\item 
The main error is in the value for $m_b - m_c$; its $\sim 1\%$ 
uncertainty stated in eq.(\ref{MBMCNUM}) translates into a $\sim 5 
\%$ one for 
$\Gamma _{SL}(B)$ and thus $\sim 3\%$ for $|V(cb)|$. 
\item 
The remaining separate sensitivity to $m_b$ generates a 
$\sim 1\%$ error. 
\item 
The first two points can be expressed through a simple scaling law  
\cite{URIROME} 
\begin{equation} 
\delta |V(cb)|_{m_b - m_c,m_b} \simeq 
\left( 1 - 0.012 \cdot \frac{\aver{(\vec p_b)^2}_B - 
0.4\, ({\rm GeV})^2}{0.1\, ({\rm GeV})^2} \right) \cdot 
\left( 1 - 0.006 \cdot \frac{\delta m_b}{30\, {\rm MeV}}\right) 
\label{ERRORIN} 
\end{equation} 
keeping in mind that at present the main uncertainty on 
$m_b - m_c$ originates in the value of 
$\aver{(\vec p_b)^2}_B$. 
\item 
Finally, one has to allow for a $\sim 1\%$ error in $|V(cb)|$ due to 
the not completely known higher order perturbative 
corrections. 
\end{enumerate} 
The purpose of this `anatomy' is not to claim that the present 
theoretical error cannot exceed 5 \% by a single iota, but to elucidate 
the bases for the estimate -- and to indicate {\em how it can be 
improved 
in the future}: 
\begin{enumerate}
\item 
The value of $m_b - m_c$ can be determined also from 
a detailed study of the shape of 
the lepton spectra in $B \ra l \nu +X_c$ 
\cite{PRL,MOMENTS}; a precision of 
$\delta (m_b - m_c) \sim 0.5 \%$ seems to be achievable 
generating $\delta |V(cb)| \sim 1\%$. Measuring the mass difference 
in this way would also avoid one assumption inherent in 
eq.(\ref{MBMCNUM}), namely that charm is sufficiently heavy, 
at least in this instance, to make an expansion in $1/m_c$ 
numerically reliable. 
\item 
Likewise one will be able to extract $m_b$ from the lepton spectra 
well enough such that $\delta |V(cb)| \sim 0.5 \%$ from this source 
alone. 
\item 
A full calculation of the $\as ^2$ corrections beyond the BLM 
prescription appears to be a feasible, though technically 
non-trivial project; the remaining perturbative error would be 
reflected in $\delta |V(cb)| \sim 0.5\%$. 
\item 
Altogether one can state as an ambitious, though feasible 
expectation for the midterm future: 
\begin{equation} 
\delta |V(cb)| \sim 2\% 
\label{VCBFUTURE} 
\end{equation} 
\end{enumerate} 
\subsubsection{$|V(cb)|$ from $\bar B \ra l \nu D^*$ at 
Zero-Recoil} 
The exclusive channel $\bar B \ra l \nu D^*$ provides us with an 
intriguing opportunity to extract $|V(cb)|$ by adopting the following  
strategy \cite{SVNW,IW}: 
\begin{enumerate}
\item 
One measures the rate for $\bar B \ra l \nu D^*$ as a function of the 
momentum transfer, extrapolates to the kinematical point 
of zero-recoil for the $D^*$ and extracts 
$|F_{\bar B \ra D^*}(0)V(cb)|$, where $F_{\bar B \ra D^*}(0)$ denotes 
the form factor at zero-recoil. The present world average yields 
\cite{CASSEL}  
\begin{equation} 
|F_{\bar B \ra D^*}(0)V(cb)| = 0.0339 \pm 0.0014 
\label{WAFZERO} 
\end{equation} 
\item 
One then has to calculate the size of the formfactor. Asymptotically, 
i.e. for $m_b$, $m_c \gg \mu _{had}$  
$F_{\bar B \ra D^*}(0) =1$ holds as a consequence of the heavy quark 
symmetry. For finite values one then has 
\cite{SVNW,LUKE}: 
\begin{equation} 
F_{\bar B \ra D^*}(0) =1 + {\cal O}\left( \frac{\as}{\pi}\right) 
+ {\cal O}\left( \frac{1}{m_c^2}, \frac{1}{m_bm_c}, 
\frac{1}{m_b^2}\right) \, , 
\label{FZERO}  
\end{equation} 
i.e. perturbative as well as non-perturbative corrections will 
drive the form factor away from unity. 
Originally it had been claimed that $F_{\bar B \ra D^*}(0) = 
0.98 \pm 0.02$ holds. The perturbative corrections are indeed 
small (though care has to be applied to their treatment); however 
the leading non-perturbative corrections should be given by 
$(\mu /m_c)^2 \sim {\cal O}(10\% )$  rather than by 
$(\mu /m_b)^2 \sim {\cal O}(2\% )$. 
\end{enumerate}
This issue can be addressed not only within HQET \cite{MANNEL}, 
but also 
through a judicious application of Heavy Quark Expansions 
to {\em inclusive} semileptonic rates. As discussed in 
\cite{OPTICAL}
\footnote{The necessary theoretical tools go beyond those 
described in Sect.2.} one can derive SV sum rules: from QCD one 
calculates the {\em nth moments} of certain transition rates 
$\bar B \ra l \nu +X_c$ as they are produced by specified weak 
currents in the SV limit and -- invoking quark-hadron duality -- 
equates them with the same moments of observable semileptonic 
rates. In particular 
one considers the case of axialvector currents which produce 
$D^*$ and its higher excitations: 
\begin{equation} 
\Gamma ^{(n)}\left( \bar B \ra l \nu ({\rm quarks \, \& \, 
gluons})_{A\times A}\right) = 
\Gamma ^{(n)}\left( \bar B \ra l \nu (D^*+{\rm excitations})\right) 
\label{SVSR} 
\end{equation}
Invoking the positivity of transition rates one can derive 
bounds on individual exclusive rates although a priori it is not 
guaranteed that such bounds are useful phenomenologically. In this 
instance, however, it turns out to be useful; this is not completely 
unexpected since even in the `extended' SV limit \cite{FIVE} a small 
number of channels dominates the inclusive transition: 
$$  
\xi _A(\mu ) - |F_{\bar B \ra D^*}(0)|^2 = 
\frac{1}{3} \frac{\aver{\mu _G^2}_B}{m_c^2} + 
\frac{\aver{(\vec p_b)^2}_B - \aver{\mu _G^2}_B} {4} 
\cdot \left( \frac{1}{m_c^2} + \frac{1}{m_b^2} + 
\frac{2}{3m_c m_b}\right) + 
$$ 
\begin{equation} 
+ \sum _{\epsilon _f < \mu } |F_{\bar B \ra {\rm f}}|^2 
+ {\cal O}(1/m_c^3, ...) 
\label{BOUND2} 
\end{equation} 
The $F_{\bar B \ra {\rm f}}$ represent the form factors for those 
charm excitations beyond the $D^*$ that are produced by the 
axialvector current with a mass $M_f = M_{D^*} + 
\epsilon _f$; $\xi _A(\mu )$ denotes a perturbative 
renormalization factor depending on $\mu$, the scale 
separating the long and short distance domains that 
was introduced in eq.(\ref{OPE}); a detailed 
discussion of this point can be found in ref. \cite{URIROME}.  
With the possible exception of corrections of order $1/m_Q^3$ all 
terms on the right-hand side of eq.(\ref{BOUND2}) are positive, see 
eq.(\ref{BOUND1}); thus 
a model independent {\em upper} bound can be placed on 
$F_{\bar B \ra D^*}(0)$ using 
$\aver{(\vec p_b)^2}_B \geq \aver{\mu _G^2}_B$ established by 
another SV sum rule. 
The bound can be further strengthened by using the 
QCD sum rule result from eq.(\ref{BALLBRAUN}). Finally one can 
make an educated estimate on the contributions from the higher 
excitations. The results are given in Table \ref{RESULTSFZERO};  
we see that indeed the deviation from unity is closer to 
10 \%  than 2 \% , as suspected!   

\begin{table}
\centering
\caption{ \it Estimates on the size of $F_{\bar B \ra D^*}(0)$.
}
\vskip 0.1 in
\begin{tabular}{|l|c|c|c|} \hline
          &  $\aver{(\vec p_b)^2}_B \geq \aver{\mu _G^2}_B$ &
 $\aver{(\vec p_b)^2}_B \simeq 0.5\; ({\rm GeV})^2$  
& Estimate  \\
\hline
\hline
 $F_{\bar B \ra D^*}(0)$  =      & $< 0.94$ & $< 0.92$ 
& $0.90 \pm 0.03$ \\
\hline
\end{tabular}
\label{RESULTSFZERO}
\end{table}

 Using the estimate for the formfactor one then infers from 
eq.(\ref{WAFZERO}: 
\begin{equation} 
|V(cb)|_{excl} = 0.0377 \pm 0.0016_{exp} \pm 
0.002_{theor} 
\label{VCBEXCL} 
\end{equation} 
which is consistent with the inclusive value, eq.(\ref{VCBIN}). 
The agreement would have been much more iffy if 
$F_{\bar B \ra D^*}(0) = 0.98 \pm 0.02$ were to hold. 

Some comments on the estimate of the theoretical uncertainty: 
\begin{itemize} 
\item 
The estimate of $\sum _{\epsilon _f \leq \mu}
|F_{\bar B \ra {\rm f}}|^2$ is 
certainly {\em not} above reasonable suspicion. 
\item 
Corrections of order $1/m_Q^3$ of which the leading term is 
presumably controlled by $1/m_c^3$ are not known in both 
magnitude and {\em sign}. They could modify the inequality. 
\item 
I am not optimistic that we can significantly reduce the theoretical uncertainty 
here -- unlike in the inclusive analysis. 
\end{itemize} 

\subsection{Extracting $|V(ub)|$} 
CLEO has seen the exclusive charmless channels $B \ra l \nu \pi \, , 
\; l \nu \rho$ \cite{CASSEL}; to determine $|V(ub)|$ there, one has to 
rely on models and I will not comment on them. The first evidence 
for $|V(ub)/V(cb)| \neq 0$ came from observing leptons with 
energies beyond the kinematical limit for $\bar B \ra l \nu X_c$. 
The quantitative analysis at present still requires some model 
elements although their weight will be reduced in the future. 

The total width $\Gamma (\bar B \ra l \nu X_u)$, on the other hand, 
can be expressed reliably in terms of $|V(ub)|$, 
$\aver{(\vec p_b)^2}_B$ and $m_b(1\, {\rm GeV})$ as discussed 
before; the last quantity is known from 
$\Upsilon$ spectroscopy. Once $\aver{(\vec p_b)^2}_B$ has been 
determined from the lepton spectrum in $\bar B \ra l \nu X_c$, 
one can extract $|V(ub)|$ reliably -- if 
$\Gamma (\bar B \ra l \nu X_u)$ can be separated out and 
measured. Taking the recent ALEPH findings 
$\Gamma (\bar B \ra l \nu X_u)/\Gamma (\bar B \ra l \nu X_c) 
= 0.016 \pm 0.004 \pm 0.004$ at face value one 
obtains \cite{URIROME} 
$|V(ub)/V(cb)| \simeq 0.098 \pm 0.013 \pm 0.013$ where the 
theoretical uncertainty is considerably smaller than the two experimental ones stated. 

\section{Lifetimes of Heavy-Flavour Hadrons}
\subsection{Generalities} 
The lifetimes of weakly decaying charm and beauty hadrons can 
of course be measured accurately without theoretical input. There 
is also no apparent qualitative disaster in the pattern observed since 
\begin{equation} 
\left| \frac{\tau (B^-)}{\tau (B_d)} -1\right| \ll 
\frac{\tau (D^+)}{\tau (D^0)} -1\; \; , \; \; 
1 - \frac{\tau (\Lambda _b)}{\tau (B_d)} >  
1 - \frac{\tau (\Lambda _c)}{\tau (D^0)} \, , 
\label{PATTERN} 
\end{equation} 
i.e., the relative difference in the $B^-$-$B_d$ lifetimes is 
considerably smaller than for the $D^+$-$D^0$ case and  
the $\Lambda _Q$ are shorter lived than the $P_Q$ with the 
effect much more pronounced in the charm than in the beauty 
sector. 

Yet the heavy quark expansions should be applicable in a 
quantitative way, at least to beauty lifetimes. Inversely a 
failure in describing these inclusive quantities would be 
quite instructive -- even if not welcome -- regarding our 
theoretical control over QCD. Yet such a failure which could be 
caused by a violation of {\em local} quark-hadron duality 
to be defined later does {\em not} prejudge our 
ability to treat semileptonic decays through a heavy-quark 
expansion. 

There is an intriguing pattern in how the lifetimes evolve and 
get differentiated order by order in $1/m_Q$; I sketch it here 
for the widths of charged and neutral pseudoscalar mesons and the 
lowest baryons: 
\begin{eqnarray} 
\Gamma (\Lambda _Q) &=& \Gamma (P^0_Q) =\Gamma (P^{\pm}_Q) 
+ {\cal O}(1/m_Q) \\ 
\Gamma (\Lambda _Q) &=& \Gamma (P^0_Q) =\Gamma (P^{\pm}_Q) 
+ {\cal O}(1/m^2_Q) \\ 
\Gamma (\Lambda _Q) &>& \Gamma (P^0_Q) \simeq \Gamma 
(P^{\pm}_Q) 
+ {\cal O}(1/m^3_Q) \\
\Gamma (\Lambda _Q) &>& \Gamma (P^0_Q) >\Gamma (P^{\pm}_Q) 
+ {\cal O}(1/m^4_Q) 
\label{WIDTHPATTERN} 
\end{eqnarray} 

\subsection{The Lifetimes of Charm Hadrons -- Predictions 
without Guarantees} 
With an expansion parameter as large as $\mu /m_c \sim 0.4$ 
one can hope for a $1/m_c$ expansion to provide us with at best a 
semi-quantitative description of charm lifetimes. 

In Table \ref{TABLE10} I juxtapose the data with the 
theoretical expectations obtained from the 
heavy quark expansion described
above. The numbers for baryon lifetimes are based on 
{\em quark model}  
evaluations of the four-fermion expectation values; this is 
indicated by an asterisk.   
Details can be found in \cite{BELLINI}.  
\begin{table}[t] 
\caption{QCD Predictions for Charm Lifetime Ratios  
\label{TABLE10}}
\vspace{0.4cm}
\begin{center}   
\begin{tabular} {|l|l|l|l|}
\hline
Observable &QCD Expectations ($1/m_c$ expansion) & Ref. &
Data from \cite{BELLINI} \\ 
\hline 
\hline 
$\tau (D^+)/\tau (D^0)$ & $\sim 2 \; \; \; $ 
[for $f_D \simeq 200$ MeV] & \cite{MIRAGE} & $2.547 \pm 0.043$ 
\\ 
&(mainly due to {\em destructive} interference)& & \\ 
\hline 
$\tau (D_s)/\tau (D^0)$ &$1\pm$ few $\times 0.01$   
& \cite{DS} &  $ 1.12\pm 0.04$ \\
\hline 
$\tau (\Lambda _c)/\tau (D^0)$&$\sim 0.5 ^*$  & \cite{MARBELLA} 
& 
$0.51\pm 0.05$\\
\hline 
$\tau (\Xi ^+ _c)/\tau (\Lambda _c)$&$\sim 1.3 ^*$ & 
\cite{MARBELLA} &  
$1.75\pm 0.36$\\
\hline
$\tau (\Xi ^+ _c)/\tau (\Xi ^0 _c)$&$\sim 2.8 ^*$ & \cite{MARBELLA} 
& 
$3.57\pm 0.91$\\
\hline 
$\tau (\Xi ^+ _c)/\tau (\Omega _c)$&$\sim 4 ^*$ & \cite{MARBELLA} 
&  
$3.9 \pm 1.7$\\
\hline
\end{tabular}
\end{center} 
\end{table} 
 
The agreement between the expectations and the data, within the 
uncertainties,  is respectable or even 
remarkable considering the large theoretical expansion parameter 
and the fact that the lifetimes for the apparently shortest-lived 
hadron -- $\Omega _c$ -- and for the longest-lived one -- 
$D^+$ -- differ by an order of magnitude! Of course the experimental 
uncertainties in $\tau (\Xi _c)$ and $\tau (\Omega _c)$ are still 
large; the present agreement could fade away 
-- or even evaporate -- 
with the advent of more accurate data. Yet at present I conclude: 
\begin{itemize} 
\item The observed difference in $\tau (D^0)$ vs. $\tau (D^+)$ is 
understood as due mainly (though not exclusively) to a destructive 
interference in $\Gamma _{NL}(D^+)$ arising in order $1/m_c^3$. This 
is 
{\em not} contradicted by the data showing $BR_{SL}(D^+)\simeq 17$ 
\%. 
For the corrections of order $1/m_c^2$ reduce the number obtained 
in the {\em naive} spectator model 
-- $BR_{SL}(D)\simeq BR_{SL}(c)$ -- from 
around 
16\% down to around 9\% \cite{BUV}! 

\item The observed near-equality of $\tau (D^0)$ and $\tau (D_s)$ 
provides us with strong, though circumstantial evidence for the reduced weight 
of WA. It puts a severe bound on the size of the 
non-factorizable parts in the expectation values of 
the four-fermion operators, as given in \cite{DS}.  

\item The lifetimes of the charm baryons reflect the interplay of 
destructive as well as constructive PI and WS intervening in order 
$1/m_c^3$ \cite{BARYONS1,BARYONS2,BARYONS3}
\footnote{The $\Omega _c$ width gets differentiated relative to the 
$\Lambda _c$ already in order $1/m_c^2$ since its expectation 
value for the chromomagnetic operator does not vanish, see 
\cite{BELLINI}}: 
\begin{eqnarray} 
\Gamma (\Lambda _c^+)&=& \Gamma _{decay}(\Lambda _c^+) +
\Gamma _{WS}(\Lambda _c^+) -
|\Gamma _{PI,-}(\Lambda _c)|   \\ 
\Gamma (\Xi _c^0)&=& \Gamma _{decay}(\Xi _c^0) +
\Gamma _{WS}(\Xi _c^0) +
|\Gamma _{PI,+}(\Xi _c^0)|   \\ 
\Gamma (\Xi _c^+)&=&\Gamma _{decay}(\Xi _c^+) +
|\Gamma _{PI,+}(\Xi _c^+)| -
|\Gamma _{PI,-}(\Xi _c^+)|   \\ 
\Gamma (\Omega _c) &=& \Gamma _{decay}(\Omega _c) + 
|\Gamma _{PI,+}(\Omega _c)| 
\label{CHARMBARYONS} 
\end{eqnarray} 
with both quantities on the right-hand-side of the last equation  
differing from the corresponding ones for $\Lambda _c$ 
or $\Xi _c$ decays \cite{BELLINI}. 
On rather general grounds one concludes:
\begin{equation} 
\tau (\Xi _c^0) < \tau (\Xi _c^+)\; , \; \; \; 
\tau (\Xi _c^0) < \tau (\Lambda _c^+) 
\end{equation} 
To go beyond this qualitative prediction one has to 
evaluate the expectation values of the various 
four-fermion operators. No model-independant manner is 
known for doing that for baryons; we do not even have a concept like 
factorization allowing us to lump our ignorence into a single 
quantity. Instead we have to rely on quark model 
computations and thus have to be prepared for 
additional very sizeable theoretical uncertainties. 

\item The $\Omega _c$ naturally emerges as the shortest-lived 
charm 
hadron due to spin-spin interactions between the decaying $c$ 
quark and the spin-one $ss$ di-quark system. 

\end{itemize}  

\noindent Finally one should note that the ratios 
$BR_{SL}(\Xi _c)/BR_{SL}(D^0)$ and 
$BR_{SL}(\Omega _c)/BR_{SL}(D^0)$ will {\em not} 
reflect 
their lifetime ratios; for $\Gamma _{SL}(\Xi _c)$ and 
$\Gamma _{SL}(\Omega _c)$ get significantly enhanced relative to 
$\Gamma _{SL}(D^0)$ in order $1/m_c^3$ due to {\em constructive} 
PI in 
$\Gamma _{SL}(\Xi _c, \Omega _c)$ among the $s$ quarks 
\cite{VOLOSHIN2}. Thus $\Omega _c$ 
-- despite its short lifetime -- could well exhibit a larger 
semileptonic branching ratio than $D^0$! 

\subsection{The Lifetimes of Beauty Hadrons -- 
Predictions Without 
Plausible Deniability}
Most of the obvious theoretical caveats one can express about 
charm lifetimes 
cannot be used as excuses for failures in beauty decays. Due to 
$m_b \gg m_c > \mu $ the heavy quark expansion would 
be expected to yield fairly reliable predictions on lifetime ratios 
among beauty hadrons. 
Since we do better than expected for charm lifetimes, one feels 
doubly confident about making predictions for the lifetimes 
of beauty hadrons. 
The actual computations proceed in close analogy to the charm case;  
details can be found in \cite{BELLINI}.  
The $B_d - B^-$ lifetime difference is again driven mainly by 
destructive PI, namely in the $b\ra c \bar ud$ channel; similarly,  
$\tau (\Lambda _b)$ is reduced relative to $\tau (B_d)$ by 
WS winning out over destructive PI in $b\ra c \bar ud$: 
$$\Gamma (B_d) \simeq \Gamma _{decay}(B_d) \; , \; 
\Gamma (\Lambda _b) \simeq \Gamma _{decay}(\Lambda _b) 
+ \Gamma _{WS}(\Lambda _b) - |\Gamma _{PI,-}(\Lambda _b) |
$$
In Table \ref{TABLE20} 
I list the world averages of {\em published} data 
together with the predictions. 
\begin{table}[t]
\caption{QCD Predictions for Beauty Lifetimes  
\label{TABLE20}}  
\begin{center}
\begin{tabular} {|l|l|l|l|}
\hline
Observable &QCD Expectations ($1/m_b$ expansion)& Ref. &
Data from \cite{BELLINI}\\ 
\hline 
\hline 
$\tau (B^-)/\tau (B_d)$ & $1+
0.05(f_B/200\, {\rm MeV} )^2
[1\pm {\cal O}(30\%)]>1$ & \cite{MIRAGE} & $1.04 \pm 0.04$ \\
&(mainly due to {\em destructive} interference)&  & \\ 
\hline 
$\bar \tau (B_s)/\tau (B_d)$ &$1\pm {\cal O}(0.01)$ & 
\cite{STONE2}  
&  $ 0.97\pm 0.05$ \\
\hline 
$\tau (\Lambda _b)/\tau (B_d)$&$\simeq 0.9 ^*$ & \cite{STONE2} & 
$0.77\pm 0.05$ \\
\hline 
\end{tabular}
\end{center} 
\end{table} 
First a few short comments for orientation: 
\begin {itemize} 
\item These are predictions in the old-fashioned 
sense, i.e. they were made before data (or data of comparable 
sensitivity) became available. 

\item As far as the meson lifetimes are concerned, 
data and predictions are completely and {\em non-trivially} 
consistent. 

\item The average $B_s$ lifetime, i.e. $\bar \tau (B_s) = 
[\tau (B_{s,{\rm long}}) + \tau (B_{s,{\rm short}})]/2$, 
as measured in $B_s \ra l \nu D_s^{(*)}$, is predicted to be 
practically idential to $\tau (B_d)$. 

\item The largest lifetime difference among beauty {\rm mesons} is 
expected to occur due to $B_s - \bar B_s$ oscillations. One predicts 
\cite{BSBS}: 
\begin{equation} 
\frac{\Delta \Gamma (B_s)}{\bar \Gamma (B_s)} \equiv 
\frac{\Gamma (B_{s,{\rm short}}) - \Gamma (B_{s,{\rm long }})}
{\bar \Gamma (B_s)} \simeq 0.18 \cdot 
\frac{(f_{B_s})^2}{(200 \, {\rm MeV})^2} 
\label{BSBSBAR}
\end{equation} 
\item The prediction on $\tau (\Lambda _b)/\tau (B_d)$ seems to be 
in conflict with the data. 
\end{itemize} 
Next I give a 
more detailed evaluation of these comparisons. 

{\bf (A) $\tau (B^-)$ vs. $\tau (B_d)$:} 

\noindent The prediction given above that the $B^-$ lifetime 
exceeds that of $B_d$ by a few percent involves assuming 
factorization to hold at a low scale $\mu _{had} \ll m_Q$. That 
has been criticized in ref.\cite{NS} where it was argued that 
neither $\tau (B^-)/\tau (B_d) <1$ nor $\tau (B^-)/\tau (B_d) 
\geq1.2$ 
would be surprising due to a failure of the factorization 
approximation {\em at any scale}. 

\noindent It is conceivable that factorization might provide a poor 
approximation for the expectation values of these four-quark 
operators -- at a significant theoretical price:  
\begin{itemize} 
\item 
The successful treatment of $\tau (D^+)$ vs. $\tau (D^0)$ vs. 
$\tau (D_s)$ was based on the factorization approximation at a 
low scale. Of course, these successes might be a mere coincidence. 
\item 
For $\tau (B^-)$ to exceed $\tau (B_d)$ by 20\% or more the 
nonfactorizable contributions have to be of a magnitude that 
-- if true -- would expose serious limitations in the 
analytical evaluations of weak matrix elements. 
\item 
Destructive interference as the main motor of a 
$B^-$-$B_d$ lifetime difference can occur only in 
$B^- \ra c \bar u d \bar u$ transitions. Since those make up no 
more than about half of all $B$ decays, a 20\% lifetime 
difference would require a $\sim$ 40\% destructive interference 
in $B^- \ra c \bar u d \bar u$ -- again an amazingly huge effect. 
\item 
The possible size of nonfactorizable contributions has been studied 
in a detailed way in ref.\cite{WA,DS} (although the reader of 
ref.\cite{NS} would not realize that). It was shown there that 
the relevant expectation values of the four-quark operators 
can be determined from comparing the lepton spectra in 
$D^0 \ra l \nu X$, $D^+ \ra l \nu X$ and $D_s \ra l \nu X$ or 
in $B_d \ra l \nu X$ and $B^- \ra l \nu X$ decays. 
\item 
{\em If} factorization passed these tests while 
$\tau (B^-)/\tau (B_d) <1$ or $\tau (B^-)/\tau (B_d) \geq1.2$ 
were observed, we had to infer that {\em local} quark-hadron 
duality did not hold 
to a sufficient degree in nonleptonic $B$ decays! 
\item 
The concept of quark-hadron duality which is essential to applying 
QCD is not always clearly defined. For our purposes it can be best 
illustrated through an analysis of the `classical' reaction 
$e^+ e^- \ra \; had$. Unlike in the case of 
deep inelastic lepton-nucleon scattering where the relevant 
large parameters are momenta from the {\em space-}like region, 
the relevant momenta here and in heavy-flavour decays are 
{\em time-}like. The transition amplitude under study thus 
contains singularities in the physical region, namely poles for 
resonances and cuts signaling particle production. This has been  
discussed explicitely for $e^+e^-$ annihilation near 
$E_{c.m.} \sim 4$ GeV  \cite{POGGIO}. On the {\em real} $E_{c.m.}$ 
axis there are two poles describing the $\psi$ and 
$\psi ^{\prime}$ resonances and there is a cut reflecting the 
production of open-charm hadrons. The cross section can 
be computed in QCD through an operator product expansion along 
the {\em imaginary} $E_{c.m.}$ axis. A dispersion relation is then 
used to continue the result into the physical regime; this means, 
however, that only `smeared' transition rates can be predicted, 
i.e. transition rates averaged over some finite energy range 
$\Delta E$: 
\begin{equation} 
\aver{\sigma (e^+e^- \ra {\rm had}; E_{c.m.})} 
\equiv \frac{1}{\Delta E}\int _{E_{c.m.}-\Delta E}^{E_{c.m.}+\Delta E}
d\tilde E \sigma (e^+e^- \ra {\rm had}; \tilde E)
\label{SMEARED} 
\end{equation} 
Equating the quantity thus calculated with the corresponding 
observed one constitutes the assumption of (global) duality. 
It was advocated in 
\cite{POGGIO} to use $\Delta E \simeq \mu  
\sim 0.5 - 1$ GeV. If the cross section happens to be a smooth 
function of $\tilde E$ -- as it happens far away from 
any production thresholds -- then one can effectively take the 
limit $\Delta E \ra 0$ to predict $\sigma (e^+e^- \ra \; had)$ for 
a {\em fixed} energy $E_{c.m.}$. This scenario is referred to as 
{\em local} duality and clearly represents a stronger assumption 
than global duality. When one describes semileptonic decays, then 
one deals with {\em smeared} quantities since integration over 
the neutrino momenta is understood; assuming global duality then 
suffices. In nonleptonic decays on the other hand such smearing is 
not guaranteed, there could be unforeseen singularities in the 
$qq\bar q \bar q$ matrix elements and in general one has to invoke 
local duality for equating the results of the heavy quark expansion 
with observable rates. 

\end{itemize} 

{\bf (B) $\tau (B_s)$ vs. $\tau (B_d)$:} 

\noindent There is general agreement that the heavy quark 
expansion predicts that the average $B_s$ lifetime as measured 
in semileptonic decay modes and the $B_d$ lifetime practically 
coincide. 

{\bf (C) $\tau (\Lambda _b)$ vs. $\tau (B_d)$:} 

\noindent The experimental situation has not been settled yet.
Let me cite here the CDF results from the full run 1 data sample 
of 110 $pb^{-1}$: 
\begin{eqnarray} 
\tau (B_d) &=& 1.52 \pm 0.06 \; psec \\ 
\tau (\Lambda _b) &=& 1.32 \pm 0.15 \pm 0.07 \; psec \\ 
\frac{\tau (\Lambda _b)}{\tau (B_d)} &=& 0.87 \pm 0.10 \pm 0.05 
\label{CDF} 
\end{eqnarray}
While this ratio is quite consistent with the stated world average, it 
would also satisfy the theoretical prediction. 

\noindent The difference between 
$\aver{\tau (\Lambda _b)/\tau (B_d)}_{exp.} \simeq 0.77$ and 
$\tau (\Lambda _b)/\tau (B_d)|_{theor} \simeq 0.9$ represents a 
large discrepancy. For once one has established -- as we have -- 
that $\tau (\Lambda _b)$ and $\tau (B_d)$ have to coincide for 
$m_b \ra \infty$, then the predictions really concern the 
deviation from unity; finding a $\sim 23$ \% deviation when one 
around 10 \% was predicted amounts to an error of about  
200 \%! 

\noindent (iii) A failure of that proportion cannot be rectified 
unless one adopts a new paradigm in evaluating baryonic expectation 
values. 
Two recent papers \cite{BOOST,NS} have re-analyzed the relevant 
{\em quark model} calculations and found:    
\begin{equation}
\tau (\Lambda _b)/\tau (B_d) \equiv 1 - {\rm DEV}\, , \; 
{\rm DEV} \sim 0.03 \div 0.12 
\label{DEVEST}
\end{equation}  
i.e., indeed there are large theoretical uncertainties in DEV since the 
baryon lifetimes reflect the interplay of several contributions of 
different signs in addition to the quark decay expression, namely 
from $WS$ and destructive as well as constructive $PI$: 
\begin{eqnarray}
\Gamma (\Lambda _b) &=& \Gamma _{decay}(\Lambda _b) 
+ \Gamma _{WS}(\Lambda _b) - 
|\Gamma _{PI,-}(\Lambda _b, b \ra c \bar u d)| \\ 
\Gamma (\Xi _b^0) &=& \Gamma _{decay}(\Xi _b) 
+ \Gamma _{WS}(\Xi _b) - 
|\Gamma _{PI,-}(\Xi _b, b \ra c \bar c s)| \\ 
\Gamma (\Xi _b^-) &=& \Gamma _{decay}(\Xi _b)  - 
|\Gamma _{PI,-}(\Xi _b, b \ra c \bar c s)|  - 
|\Gamma _{PI,-}(\Xi _b, b \ra c \bar u d)|  
\end{eqnarray} 
Yet 
one cannot boost the size of DEV much beyond the 10 \% level. To 
achieve the 
latter one had to go {\em beyond} a description of baryons in terms 
of three 
valence quarks only. A similar conclusion has been reached by 
the authors of ref.\cite{BARI} who analyzed the relevant 
baryonic matrix elements through QCDJsum rules. 

\subsection{A Radical Phenomenological Proposal}
In a recent paper \cite{ALTARELLI} it was argued  
that the widths of heavy-flavour 
hadrons scale with the fifth power of their mass -- $M_{H_Q}$ -- 
rather than the heavy quark mass $m_Q$. This is inconsistent with 
the heavy quark expansion based on the operator product expansion:  
it introduces corrections of order $1/m_Q$ in a prominent way:
\begin{equation} 
\Gamma (H_Q) \propto G_F^2 M^5_{H_Q} = 
G_F^2 ( m_Q + \bar \Lambda + ...)^5 = 
G_F^2 m_Q^5 \left( 1 + 5 \cdot \frac{\bar \Lambda}{m_Q} + ...\right) 
\label{ANSATZ} 
\end{equation} 
However in the spirit of the time-honoured advice 
of `Peccate Fortiter' it was suggested that local duality 
does not hold in nonleptonic decays of beauty and charm 
hadrons.  

The recipe leads to 
\footnote{This suggestion was actually first made in 
\cite{MIRAGE} when the experimental evidence for it was quite 
marginal before the authors realized it was inconsistent with 
the operator product expansion.}: 
\begin{equation} 
\frac{\tau (\Lambda _b)}{\tau (B_d)} \simeq 
\left( \frac{M_{B_d}}{M_{\Lambda _b}}\right) ^5 \simeq 
0.77 \pm 0.05 
\end{equation} 
Likewise one finds for the average $B_s$ lifetime: 
\begin{equation} 
\frac{\bar \tau (B_s)}{\tau (B_d)} \simeq 
0.93 \pm 0.03 \; ,  
\end{equation} 
which is a significantly smaller ratio than predicted by the 
$1/m_Q$ expansion, yet quite consistent with present 
measurements. 

When applying this prescription to charm decays one finds that 
the observed $\Lambda _c$, $\Xi _c^0$ and $\Omega _c$ lifetimes 
follow the scaling law of eq.(\ref{ANSATZ}) relative to the 
$D^0$ lifetime. However a pattern 
$$ 
\tau (D^+) \simeq \tau (D^0) > \tau (D_s) > \tau (\Xi _c^+) 
$$ 
since $M(D^+) \simeq M(D^0) < M(D_s) < M(\Xi _c^+)$ is in obvious 
conflict with the data. This has to be remedied by the a posteriori 
introduction of destructive $PI$ in $D^+$ and $\Xi _C^+$ decays and 
of {\em destructive} $WA$ in $D_s$ decays tuned as to reproduce 
the data. One should note, though, that a large overall destructive 
$PI$ contribution is not natural for $\Gamma (\Xi _c^+)$ since 
there arises also a {\em constructive} $PI$ term, see 
eqs.(\ref{CHARMBARYONS}); a $WA$ contribution to 
$\Gamma (D_s)$ that is both sizeable and destructive would be 
surprising as well. 

Measurements of the semileptonic branching ratios for 
$\Xi _c^{0,+}$ and $\Omega _c$ baryons would provide 
important constraints for this phenomenological model 
as well as for the OPE based heavy quark expansion. 

\subsection{$B_c$ Decays} 
 $B_c$ decays provide a particularly intriguing lab to probe QCD. 
They are shaped by three classes of reactions, namely the decay 
of the $b$ quark, the $c$ quark and WA between the two heavy 
constituents. 

It had been suggested that the quark masses entering the quark 
decay widths should be reduced by the binding energy of the 
$\bar bc$ bound state; this would imply that the $B_c$ lifetime 
is relatively long, namely above 1 psec with beauty decays 
dominating over charm decays. However the $1/m_Q$ expansion 
\cite{BCLAB,BENEKE} 
predicts a lifetime well below 1 psec with unfortunately charm 
decays dominating, mainly because there are no corrections of 
order $1/m_Q$. These findings are also in agreement with an 
earlier phenomenological analysis \cite{LUSIGNOLI}. It is 
curious to note that the recipe of ref.\cite{ALTARELLI} would also 
yield a short $B_c$ lifetime: 
$\tau (B_c) \sim (M_B/M_{B_c})^5 \tau (B_d) \sim 0.6$ psec for 
an expected mass value of $M_{B_c} = 6.26$ GeV. It is 
not clear what this ansatz predicts for the relative weight of $b$ and 
$c$ decays. 

A study of $B_c$ decays would thus provide us with crucial tests -- 
alas, the prospects for that to happen soon are quite gloomy 
\cite{RUECKL}! 

\section{Summary and Outlook}
Very considerable progress has been achieved in the theoretical 
description of heavy-flavour decays both of the inclusive 
variety, as mainly discussed here, and the exclusive one 
\cite{MANNEL}. 
We can treat {\em semileptonic} transitions of beauty hadrons 
with a reliability and precision that would have seemed unrealistic 
a few years ago. This can be illustrated by the values extracted 
for $|V(cb)|$ from inclusive and exclusive semileptonic $B$ 
decays: 
\begin{eqnarray} 
|V(cb)|_{incl} &=& 0.0413 \pm 0.0016_{experim} \pm 
0.002_{theor} \\
|V(cb)|_{excl} &=& 0.0377 \pm 0.0016_{experim} \pm 
0.002_{theor} 
\end{eqnarray} 
In each case the experimental and theoretical uncertainties are 
comparable and the two values are consistent with each other. 
This is quite remarkable considering that they result from 
analyses that are quite different systematically in their 
experimental as well as theoretical elements. This agreement 
came about in a non-trivial way since it is based on the 
formfactor $F_{B \ra D^*}(0)$ to be substantially smaller than 
unity. At least as far as the theoretical treatment of 
$\Gamma _{SL}(B)$ is concerned I am confident that the 
theoretical uncertainty can be reduced significantly in the 
foreseeable future. 

With respect to {\em nonleptonic} decays we can now tackle 
questions that could not be addressed before or only in an 
ambiguous way: what is the impact of $WA$; how do 
$\tau (D_s)$ and $\tau (D^0)$ or $\bar \tau (B_s)$ and 
$\tau (B_d)$ compare to each other; how does the ratio 
$\tau (P_Q^{\pm})/\tau (P_Q^0)$ scale with $m_Q$ etc. A 
failure to describe weak lifetimes will of course never 
rule out QCD; yet it will still teach us important lessons on 
QCD and our theoretical control over it. Such failures can arise 
at different layers leading to different kinds of lesson: 
\begin{itemize} 
\item 
The apparent agreement between predictions for and measurements 
of charmed baryon lifetimes might evaporate when the merciful 
imprecision of the present data is overcome. One could then 
just shrug the shoulders saying that charm baryon lifetimes 
receiving contributions from various sources with different signs 
provide a {\em numerically very unstable scenario}. 
\item 
A failure in reproducing $\tau (D_s)$ vs. $\tau (D^0)$ could be 
blamed on $m_c$ being too small 
to provide us with a reliable expansion parameter. 
\item 
The $\Lambda _b$ lifetime remaining `short' signals 
{\em at the very 
least} the need for a new paradigm in evaluating baryonic 
matrix elements. 
\item 
A failure in $\tau (B^-)$ vs. $\tau (B_d)$ vs. $\bar \tau (B_s)$ would 
cast serious doubts on factorization as useful approximation in this 
case; it would establish the short comings of {\em local} 
duality in nonleptonic beauty decays if factorization had passed 
the independant tests referred to before. 

\end{itemize}  
Of course there exists still a strong need to expand the data base: 
\begin{enumerate}
\item 
In the charm sector one wants to measure $\tau (D_s)$ with a 
2\% accuracy and $\tau (\Xi _c^+,\, \Xi _c^0,\, \Omega _c)$ with 
about 10\%. 
\item 
In the beauty sector one has to probe for percent differences in 
$\tau (B^-)$ vs. $\tau (B_d)$ vs. $\bar \tau (B_s)$. 
\item 
A dedicated effort has to be made to search for 
$\tau (B_{s, short})$ vs. $\tau (B_{s, long})$. 
\item 
One has to measure $\tau (\Lambda _b)$ with even more 
precision and study $\tau (\Xi _b^-)$ and $\tau (\Xi _b^0)$ 
as well. 
\item 
$B_c$ decays will provide an intriguing lab for the 
discriminating connoisseur! 

\end{enumerate} 

\vspace{1cm}

\noindent {\bf Acknowledgements:} I have benefited 
tremendously over the years from my collaboration with 
M. Shifman, N. Uraltsev and A. Vainshtein. I have 
thoroughly enjoyed the open and inspiring atmosphere 
at the conference and the site chosen by the organizers. 
This work was supported by the National Science Foundation under 
grant number PHY 92-13313. 
\end{document}